\documentstyle[epsfig,preprint,aps,floats]{revtex}
\tighten

\newcommand{\pow}[3]{\left(\frac{#1}{#2}\right)^{#3}}
\def\F{{\rm F}}
\def\rmd{{\rm d}}
\def\SUSY{{\rm SUSY}}

\begin{document}
\draft 
\small
\preprint{OUTP-02-09-P}
\title{Neutralino spectrum in top-down models of UHECR}
\author{\large Alejandro Ibarra\thanks{ibarra@thphys.ox.ac.uk} and 
         Ramon Toldr\`a\thanks{toldra@thphys.ox.ac.uk}}
\address{Theoretical Physics, University of Oxford,
          1 Keble Road, Oxford OX1 3NP, UK}
\maketitle

\begin{abstract}
  We calculate the cosmic ray spectrum of ultra high energy
  neutralinos that one should expect provided that the observed ultra
  high energy cosmic rays are produced by the decay of superheavy
  particles $X$, $M_X>10^{12}$~GeV, in supersymmetric models. Our
  calculation uses an extended DGLAP formalism. Forthcoming cosmic ray
  observatories should be able to detect these neutralinos.
\end{abstract}
\pacs{98.70.Sa; 14.80.-j; 95.35.+d; 13.87.Fh}

\section{Introduction}\label{intro}

Over one hundred cosmic ray events with energy higher than $4\times
10^{10}$~GeV have been detected by past and present observatories.
About 30 events have energy in the range $(1-3)\times10^{11}$~GeV
(see~\cite{NaganoWatson} for an experimental review). In top-down
production mechanisms of Ultra High Energy Cosmic Rays (UHECR) the
decay products of very massive particles $X$ with $M_X > 10^{12}$~GeV
account for these events. Particle X can be either a GUT scale
particle produced by the decay of a cosmological network of
topological defects created in a phase transition in the early
universe~\cite{Hill,td,bbv98,tdgamnu}, or a cold dark matter particle
clustered in the galactic halo whose abundance was generated during or
shortly after
inflation~\cite{chor,bkv97,BirkelSarkar,gravprod,reheatprod}.
Whatever the nature of $X$ is, it decays into partons which hadronise
and give the UHECR primaries that we observe on the Earth. Therefore,
in order to calculate the predicted spectrum and composition of UHECRs
one has to calculate the nonperturbative Fragmentation Functions (FFs)
of partons into hadrons at the energy scale $M_X$. Several approaches
have been taken to solve the fragmentation process and calculate the
spectrum of UHECRs: Montecarlo
generators~\cite{BirkelSarkar,BerezKachel01}, analytical estimates
(MLLA)~\cite{bbv98,tdgamnu,bkv97,bk98} and DGLAP
evolution~\cite{Rubin,FodorKatz,SarkarToldra,Toldra}
(see~\cite{BirkelSarkar,SarkarToldra} for a comparison of these
different approaches).

As long as supersymmetry (SUSY) is a low energy symmetry of nature,
the decay of $X$ will produce supersymmetric particles. If R-parity is
conserved, the Lightest Supersymmetric Particle (LSP) is stable and
should be among the primary cosmic rays that reach the Earth.
Theoretical motivations favour the neutralino as~the~LSP, hence we
shall concentrate on neutralinos\footnote{In some SUSY models the
  gravitino, the gluino and even the sneutrino could be the LSP. An
  exotic hadron state containing a stable gluino has been proposed to
  explain the UHECR events~\cite{ChungFarrarK}.}. In the Minimal
Supersymmetric Standard Model (MSSM) with R-parity conserved,
neutralinos are a mixture of higgsinos and neutral gauginos (bino,
$\widetilde B$, and neutral wino, $\widetilde W^0$). The precise
composition of the neutralinos in terms of the interaction eigenstates
depends on the particular SUSY scenario.  For simplicity, we shall
concentrate on the Constrained MSSM, which assumes that the soft
SUSY-breaking scalar masses, the gaugino masses and the trilinear
parameters are all universal at the GUT scale. In this scenario, it
can be shown that in the region of the parameter space where the relic
density of neutralinos is of cosmological significance, the LSP is
mainly a bino \cite{LSP}.

Weakly interacting particles like neutralinos or neutrinos have cross
sections with ordinary matter too small to be detected with present
day UHECR observatories. Baryons, nuclei and perhaps photons may
account for all the events detected so far~\cite{AveHinton}.
Forthcoming detectors will be sensitive to weakly interacting UHECRs
(see Refs.~\cite{HalzenHooper,Sigl} and references therein). It is
therefore important to obtain a good estimate of the neutralino flux
expected in top-down models.

The neutralino spectrum in top-down models has already been estimated
using Montecarlo generators by
Berezinsky\&Kachelrie\ss~\cite{BerezKachel01,BerezKachel98}. In the
Montecarlo approach a primary parton with energy $M_X/2$ produced in
the decay of $X$ initiates a parton cascade which proceeds until a
specific minimal virtuality is reached. The squarks and gluinos in the
parton shower decay into the LSP and ordinary partons once they reach
the scale $M_\SUSY$, the universal mass of squarks and gluinos. All
quark and gluons evolve down to the hadronization scale where a
phenomenological model is employed to bind them into hadrons.

The Montecarlo simulations did not include the electroweak radiation
of neutralinos by off-shell partons in the shower. In the next section
we show how the DGLAP formalism employed to calculate the neutrino,
photon and baryon spectra of
UHECR~\cite{Rubin,FodorKatz,SarkarToldra,Toldra} can be extended to
calculate the spectrum of neutralinos. We concentrate on the
electroweak radiation of neutralinos by off-shell partons to
complement the Montecarlo simulations. We find that at large $x$
electroweak radiation of neutralinos is important. The large $x$
region is the most relevant for future cosmic ray observatories since
in this region the neutralino flux is larger than the neutrino flux
and will not be swamped by the neutrino signal.

\section{Evolution equations for neutralino fragmentation functions}

The main channels of neutralino production are the hadronic decays of
$X$. Hence the flux of neutralinos is given by the fragmentation
functions from partons, $D^\chi_a(x,M_X^2)$, where $a$ is any quark
$q_k$ and squark flavour $s_k$, a gluon $g$ or a gluino $\lambda$.
The flux is proportional to the sum over $a$ of $D^\chi_a(x,M_X^2)$
each one weighted with the branching ration for the decay of $X$ into
parton $a$. The variable $0<x<1$ is the fraction of the momentum of
$a$ carried off by $\chi$. These nonperturbative functions depend on
the energy scale $\mu$; in our case $\mu=M_X$. The rate of change of
$D^\chi_a(x,\mu^2)$ with $\mu^2$ is given by the sum of two terms. The
first one takes into account the ordinary SUSY QCD branching of
partons and gives rise to the so-called
Dokshitzer--Gribov--Lipatov--Altarelli--Parisi (DGLAP)
equations~\cite{AltarelliParisi,DGL}. The second term takes into
account the emission of neutralinos in the parton shower, which stems
from the tree level weak SUSY coupling among quarks, squarks and
neutralinos. For any (anti)quark and (anti)squark we obtain, to
lowest order in the strong coupling $\alpha_S$ and the weak couplings
$\alpha^{\chi,k}_W$,
\begin{eqnarray}
  \label{eq:ModDGLAPq}
  \mu^2\partial_{\mu^2} D^\chi_{q_k}(x,\mu^2) &=&
  \frac{\alpha_S(\mu^2)}{2\pi}\left[ \sum_l P_{q_lq_k}(x) \otimes
    D^\chi_{q_l}(x,\mu^2) +\dots \right]+\frac{\alpha^{\chi,k}_W(\mu^2)}{2\pi}
  P_{\chi q_k}(x)\otimes D^\chi_\chi(x,\mu^2),\\
  \label{eq:ModDGLAPs}
 \mu^2\partial_{\mu^2} D^\chi_{s_k}(x,\mu^2) &=& 
  \frac{\alpha_S(\mu^2)}{2\pi}\left[ \sum_l P_{q_ls_k}(x) \otimes
    D^\chi_{q_l}(x,\mu^2) +\dots \right]+\frac{\alpha^{\chi,k}_W(\mu^2)}{2\pi}
  P_{\chi s_k}(x)\otimes D^\chi_\chi(x,\mu^2),
\end{eqnarray}
where the dots inside the square brackets stand for the remaining
leading order SUSY QCD terms\footnote{The DGLAP equations do not
  include soft gluon emission coherence. Coherence is important at low
  $x$. Our results will only hold for $x$ much larger than the
  position of the Gaussian peak produced by coherence, at $x_p\simeq
  \sqrt{\Lambda_{\rm QCD}/M_X}$. As pointed out in Sec.~\ref{intro} we
  are only interested in the large $x$ region. The DGLAP equations
  can be modified to include coherence, see for
  example~\cite{Rubin}.}, see
\cite{SarkarToldra,KounnasRoss,JonesLlewellyn}. For gluons and
gluinos, to lowest order in $\alpha_W$, there are no additional terms
to the usual SUSY QCD ones. The FF of neutralinos from neutralinos is
$D^\chi_\chi(x,\mu^2)=\delta(1-x)+O(\alpha_W)$. Note that the
functions $D^\chi_a(x,\mu^2)$ are of order $\alpha^{\chi,a}_W$. The
splitting functions of quarks and squarks into neutralinos are given
by $P_{\chi q_k}=1-x$ and $P_{\chi s_k}=1$.  We make use of the
convolution operator
\begin{equation}
\label{eq:Convolution}
A(x) \otimes B(x) \equiv \int^1_x \frac{\rmd z}{z}\, A(z) B(\case{x}{z}).
\end{equation}
Analogous equations to Eqs.~(\ref{eq:ModDGLAPq}--\ref{eq:ModDGLAPs})
employed to study photon structure or photon production by parton
showers can be found in \cite{DeWitt,Nicolaidis}.

It is useful to introduce the following evolution variable
\begin{equation}
 \label{eq:Tau}
 \tau \equiv \frac{1}{2\pi b}\ln\frac{\alpha_S(\mu^2_0)}{\alpha_S(\mu^2)},
\end{equation}
$b$ being the coefficient in the leading order $\beta$-function
governing the running of the strong coupling:
$\beta(\alpha_S)=-b\alpha_S^2$.

For simplicity we shall only consider gaugino production. We shall
argue later that the final results are not substantially altered if
higgsinos are included in the calculation. Also for clarity we shall
concentrate on the evolution of quark and squark singlet functions,
defined as the following sum over flavours
\begin{eqnarray}
  \label{eq:QSinglet}
D^\chi_q &\equiv& \sum_k D^\chi_{q_k}+ D^\chi_{\bar{q}_k}, \\
  \label{eq:SSinglet}
D^\chi_s &\equiv& \sum_k D^\chi_{s_k}+ D^\chi_{\bar{s}_k}.
\end{eqnarray}
The singlet functions are coupled to the gluons and gluinos by means
of the following $4\times 4$ matrix integro-differential equation
\begin{equation}
  \label{eq:SUSY4X4}
 \partial_\tau 
  \left(
    \begin{array}{l}
      D^\chi_q \\
      D^\chi_g \\
      D^\chi_s \\
      D^\chi_\lambda
    \end{array}
  \right)
  =
  \left(
    \begin{array}{cccc}
      P_{qq} & 2n_{\F}P_{gq} & P_{sq} & 2n_{\F}P_{\lambda q} \\
      P_{qg} & P_{gg} & P_{sg} & P_{\lambda g}  \\
    P_{qs} & 2n_{\F}P_{gs} & P_{ss} & 2n_{\F}P_{\lambda s} \\
   P_{q\lambda} & P_{g\lambda} & P_{s\lambda} & P_{\lambda\lambda}
    \end{array}
  \right)
  \otimes
  \left(
    \begin{array}{l}
      D^\chi_q \\
      D^\chi_g \\
      D^\chi_s \\
      D^\chi_\lambda
    \end{array}
  \right)+
  \frac{\alpha_W^\chi}{\alpha_S}
  \left(
    \begin{array}{c}
      P_{\chi q} \\
      0 \\
      P_{\chi s} \\
      0
    \end{array}
  \right),
\end{equation}
where $\alpha_W^\chi \equiv \sum_k \alpha_W^{\chi,k}$ and $n_\F$ is
the number of flavours. These will be the only relevant equations if
one assumes flavour universality in the decay of $X$. If one wishes to
study the evolution of each flavour separately then one needs the
complete set of DGLAP equations as given in
\cite{SarkarToldra,KounnasRoss,JonesLlewellyn} to which one has to add
the neutralino radiation terms on the right hand side of
Eqs.~(\ref{eq:ModDGLAPq}--\ref{eq:ModDGLAPs}).

We can write Eq.~(\ref{eq:SUSY4X4}) in the simple matrix form
\begin{equation}
  \label{eq:MatrixEq}
  \partial_\tau xD^\chi = xP \otimes  xD^\chi + xX^\chi,
\end{equation}
where $xD^\chi$ and $xX^\chi$ are $4-$vectors and $xP$ is a $4\times
4$ matrix whose elements are all the leading order SUSY QCD splitting
functions $P_{ab}$ times $x$. These functions were calculated in
\cite{KounnasRoss,JonesLlewellyn}. We have multiplied
Eq.~(\ref{eq:SUSY4X4}) by $x$ to improve numerical stability.

Equation~(\ref{eq:MatrixEq}) is a linear inhomogeneous equation. Its
associated homogeneous equation is the usual DGLAP equation for the
coupled evolution of the quark and squark singlets, gluons and
gluinos. The inhomogeneous or source term arises from the neutralino
radiation by partons. The general solution to Eq.~(\ref{eq:MatrixEq})
is the sum of the general solution to the associated homogeneous
equation plus a particular solution
\begin{equation}
  \label{eq:GeneralSolution}
  xD^\chi(x,\tau) = E(x,\tau-\tau_0) \otimes xD^\chi(x,\tau_0)+
  \int^\tau_{\tau_0} \rmd \tau' E(x,\tau-\tau') \otimes xX^\chi(x,\tau').
\end{equation}
The evolution operator $E(x,\tau)$ is the solution to the associated
homogeneous equation with the initial condition
$E(x,0)=\delta(1-x)I_4$ (see \cite{Toldra,FurmanskiPetronzio})
\begin{equation}
  \label{eq:EvolutionOperator}
  E(x,\tau)= e^{xP(x)\tau} \equiv \delta(1-x)I_4 + xP(x)\tau +
  \frac{1}{2!}xP(x)\otimes xP(x) \tau^2 + \dots
\end{equation}
We take as initial energy scale $\mu_0=M_\SUSY$, $\tau_0\equiv
\tau(\mu_0)=0$, the typical scale for all sparticle masses.

Equation~(\ref{eq:GeneralSolution}) gives the total neutralino FF as
the sum of two terms. The first one, the solution to the homogeneous
equation, is the contribution from partons that decay on-shell into
neutralinos (assuming a universal mass for all squarks and gluinos
$M_\SUSY$). This contribution was studied with Montecarlo generators
in Refs.~\cite{BerezKachel01,BerezKachel98}. The second term is the
contribution stemming from the electroweak radiation of neutralinos by
partons. From now on, we concentrate on this second term and take
\begin{equation}
  \label{eq:MainEq}
  xD^\chi(x,\tau) = 
  \int^\tau_0 \rmd \tau' E(x,\tau-\tau') \otimes xX^\chi(x,\tau').
\end{equation}
This expression is our main equation. It will allow us to estimate the 
relevance of electroweak radiative emission of neutralinos. 

\section{Numerical calculation and results}

In order to calculate numerically $xD^\chi(x,\tau)$ as given in
Eq.~(\ref{eq:MainEq}) we expand the evolution operator in Laguerre
polynomials\footnote{Laguerre polynomial expansions to study scaling
  violations in QCD were first introduced
  in~\cite{FurmanskiPetronzio}.} $L_n(x)$
\begin{eqnarray}
  \label{eq:ECoefficients}
   E(x,\tau) &=& \sum^\infty_{n=0} E_n(\tau) L_n(-\ln x), \\
   E_n(\tau) &=& \sum^4_{i=1} e^{\lambda_i\tau}\sum^n_{k=0}
   \frac{\tau^k}{k!} B^k_{i,n}.
\end{eqnarray}
The scalars $\lambda_i$ are the eigenvalues of $xP_0$, the first
coefficient in the Laguerre expansion of the matrix $xP(x)$. Energy
conservation gives $\lambda_1=0$. All other eigenvalues are negative.
The $4\times4$ matrices $B^k_{i,n}$ are given by recursive relations
that were calculated in~\cite{Toldra}. Likewise we expand the source
term in Laguerre polynomials
\begin{eqnarray}
  \label{eq:SourceTerm}
  xX^\chi(x,\tau) &=& \frac{\alpha_W^\chi(\tau)}{\alpha_S(\tau)}
  f(x), \\
  f(x) &=& \sum^\infty_{n=0} f_n L_n(-\ln x), \\
  f^T_n &=& \left( \pow{1}{2}{n+1}-
    \frac{1}{3}\pow{2}{3}{n},0,\pow{1}{2}{n+1},0\right). 
\end{eqnarray}
The Laguerre expansions for the neutralino FFs are
\begin{equation}
  \label{eq:LaguerreExp}
  xD^\chi(x,\tau) = \sum^\infty_{N=0} \tilde{C}_N^\chi(\tau) L_N(-\ln x),
\end{equation}
where $\tilde{C}_0^\chi \equiv C_0^\chi$ and $\tilde{C}_N^\chi \equiv
C_N^\chi-C_{N-1}^\chi$ for $N>0$. The $4-$vectors $C^\chi_N$ are
calculated substituting all Laguerre expansions in
Eq.~(\ref{eq:MainEq})
\begin{equation}
  \label{eq:CCoefficients}
  C_N^\chi(\tau) = \sum^N_{n=0} \sum^n_{k=0} \sum^4_{i=1} 
  \frac{B^k_{i,n} f_{N-n}}{k!} \int^\tau_0 \rmd \tau' 
  e^{\lambda_i(\tau-\tau')} (\tau-\tau')^k
  \frac{\alpha_W^\chi(\tau')}{\alpha_S(\tau')}. 
\end{equation}

\begin{figure}[bth]
  \begin{center}
    \epsfig{figure=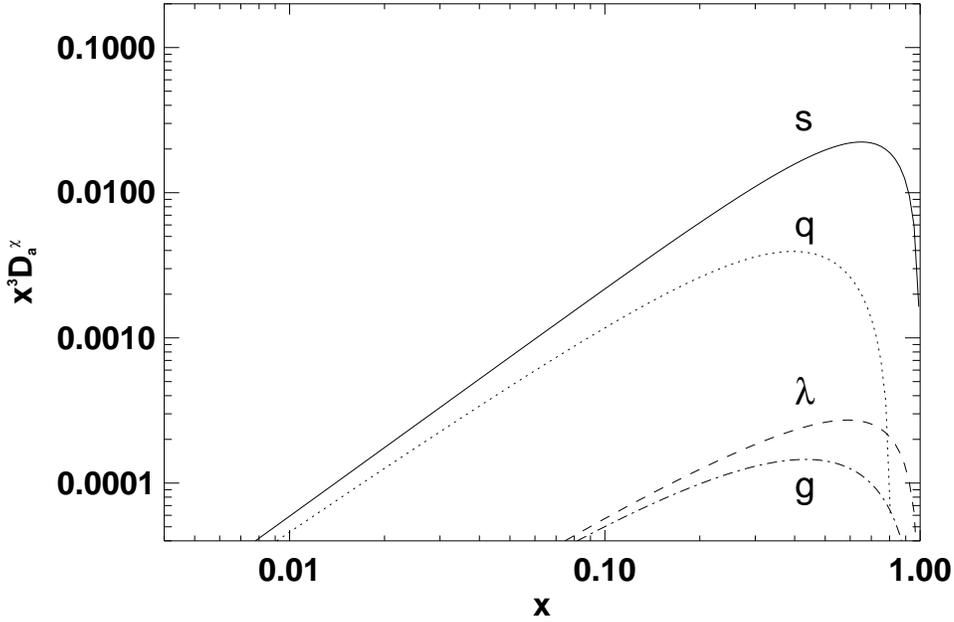,width=12cm}
    \bigskip
    \caption{$\widetilde{B}$ fragmentation functions for $M_X=10^{12}$ GeV and
      $M_\SUSY=400$~GeV. The solid line is the squark contribution,
      the dotted lines is the quark one, the dashed line is the gluino
      contribution and the dot-dashed line is the gluon one.}
     \label{fig:partons}
  \end{center}
\end{figure}

In order to present our results we multiply FFs by $x^3$ since the
measured spectrum is usually multiplied by $E^3$ to highlight its
structure. We plot in Fig.~(\ref{fig:partons}) our numerical
calculation for the bino $x^3D^{\widetilde{B}}_a(x,M^2_X)$ when
$M_X=10^{12}$~GeV and $M_\SUSY=400$~GeV. The major contribution comes
from squarks and quarks, which can radiate neutralinos at order
$\alpha_W$. The contribution from gluinos and gluons is much smaller
since they do not have order $\alpha_W$ coupling to neutralinos,
therefore to lowest order they can only generate neutralinos through
mixing with squarks and quarks, see Eq.~(\ref{eq:SUSY4X4}).
The fraction of available energy $M_X=10^{12}$~GeV carried by binos is
$\left<xD^{\widetilde{B}}_a(x,M_X)\right>=
C^{\widetilde{B}}_0(\tau(M_X))=(0.047,0.002, 0.109,0.003)$ for
$q,g,s,\lambda$, respectively (divide by $2n_\F$ to get fraction per
quark and squark flavour).  From~Eqs.~(\ref{eq:MainEq})
and~(\ref{eq:SourceTerm}) one can check that for the quark and squarks
singlets the order of magnitude is given by
$\left<xD^{\chi}\right>\sim \tau \alpha_W/\alpha_S$.

\begin{figure}[tbh]
  \begin{center}
    \epsfig{figure=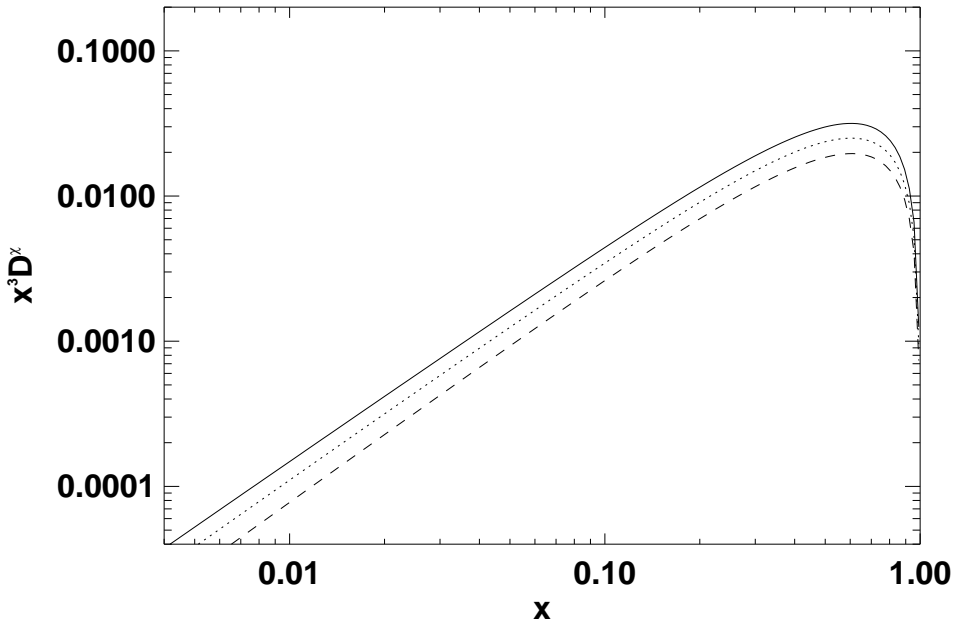,width=12cm}
    \bigskip
    \caption{$\widetilde{B}$ fragmentation functions 
      for $M_\SUSY=400$~GeV and $M_X=10^{14}$~GeV~(solid line),
      $M_X=10^{12}$~GeV~(dotted line) and $M_X=10^{10}$~GeV~(dashed line).}
     \label{fig:scaling}
  \end{center}
\end{figure}

In Fig.~(\ref{fig:scaling}) we show how $\widetilde{B}$ distributions
change for different values of $M_X$. Each curve is the sum of the
quark and squark singlets, gluon and gluino contributions. Note that
$\partial_\tau D^\chi > 0$ for all $x$ because of the inhomogeneous
term in Eq.~(\ref{eq:SUSY4X4}). If the sparticle mass scale $M_\SUSY$
is around the electroweak scale, the final result depends feebly on
$M_\SUSY$~\cite{SarkarToldra}.

The decay of $X$ will produce neutral winos as well. The
$\widetilde{B}$ and $\widetilde W^0$ curves have similar shapes, the
main difference between them being that, for a common scale $M_X$,
$\widetilde W^0$ FFs are always slightly larger than $\widetilde{B}$
FFs, since the coupling $\alpha_W^\chi$ is slightly stronger for
$\widetilde W^0$ than for $\widetilde{B}$. The wino will eventually
decay into the bino, the LSP. Its lifetime times its Lorentz factor is
much smaller than 1 kpc/$c$, therefore neutral winos produced in the
galactic halo or further will disintegrate into the LSP before
reaching the Earth. The $\widetilde{B}$ spectrum on the Earth will be
the sum of the $\widetilde{B}$ contribution produced at the decay spot
of $X$ plus the $\widetilde{B}$ contribution stemming from $\widetilde
W^0$ decay into $\widetilde{B}$.

The decay sequence of a neutral wino into a bino depends on the
details of the scenario considered. Even within the framework of the
Constrained MSSM, the freedom is still large. Nevertheless, some
benchmark points have been proposed \cite{Battaglia} for study at the
Tevatron collider, the LHC and $e^+ e^-$ colliders. These points are
consistent with different experimental constraints, as well as
cosmology, and can be regarded as generic in the whole parameter
space. We shall be only concerned with the benchmark points A-D and
G-M in Ref.~\cite{Battaglia}, where the LSP is mainly the bino and the
next-to-LSP is the neutral wino, with some admixture of higgsinos.
The two remaining points correspond to the so-called ``focus-point''
region at large scalar masses, that we shall not consider. In any
case, our analysis covers a large region of the allowed CMSSM
parameter space.

The neutral wino decays mostly into sleptons and leptons, and it is
followed by a decay of the slepton into the LSP. Therefore, only
two-body decays are relevant. Given the generic case $A\rightarrow
B+C$, we want to calculate which is the distribution of the decay
product $\rmd n_C/\rmd x$ once the distribution for the decaying
particle $\rmd n_A/\rmd x$ is known. From the phase space of the
disintegration process we obtain
\begin{equation}
  \label{eq:distribution}
  \frac{\rmd n_C(x)}{\rmd x} = \frac{m^2_A}{\sigma_2} 
  \int^{y_{max}}_{y_{min}} \frac{\rmd y}{y} \frac{\rmd n_A(y)}{\rmd y},
\end{equation}
where we have defined
\begin{eqnarray}
  \label{eq:def}
  y_{max} & \equiv &
  \mbox{min}\left[\frac{\sigma_1+\sigma_2}{2m^2_C}x,1\right], \\
  y_{min} & \equiv & \frac{\sigma_1-\sigma_2}{2m^2_C}x, \\
  \sigma_1 & \equiv & m^2_A+m^2_C-m^2_B, \\
  \sigma_2 & \equiv & \sqrt{\lambda(m^2_A,m^2_B,m^2_C)}, \\
  \lambda(a,b,c) & \equiv & a^2+b^2+c^2-2ab-2ac-2bc.
\end{eqnarray}
In addition there is the following kinematical constraint
\begin{equation}
  \label{eq:constraint}
  0 < x < \frac{2m^2_C}{\sigma_1-\sigma_2}.
\end{equation}

The LSP flux reaching the atmosphere is the sum of the
$\widetilde{B}$~contribution produced directly in the decay~of~$X$
plus the $\widetilde{B}$~contribution produced by the
decay~of~$\widetilde W^0$ on its way to the Earth. The two-body decay
of winos into binos pushes the momenta to lower values of~$x$, making
the direct $\widetilde{B}$~component dominate over the component
produced by~$\widetilde W^0$~decay, for $x>0.1$.

So far we have ignored higgsino production, so some words are now in
order. The magnitude of the gaugino and higgsino couplings to partons
are comparable, hence their FFs will be also similar. For the
benchmark points that we have analysed, higgsinos will decay on their
way to the Earth into winos and subsequently into binos. Therefore,
for the same kinematical reason as the wino case discussed above, the
total number of binos on Earth coming from higgsino decay will be a
small correction to the direct bino contribution, for $x>0.1$. For the
same reason, we have ignored binos coming from the decay of the
charginos that are also produced in the parton shower.

\begin{figure}[bth]
  \begin{center}
    \epsfig{figure=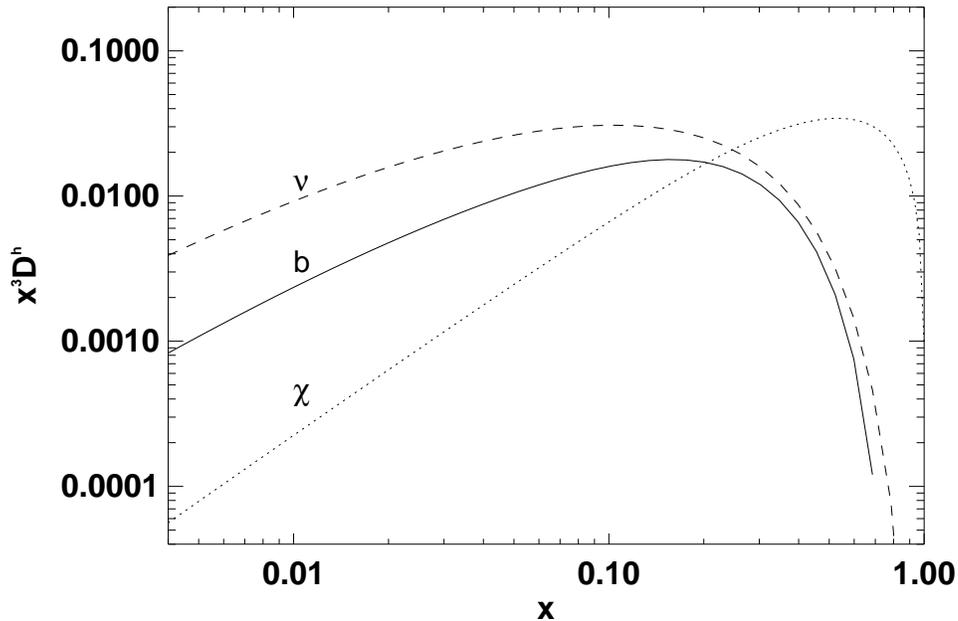,width=12cm}
    \bigskip
    \caption{Neutrino (dashed lime), baryon (solid line) and
      neutralino (dotted line) spectra expected from the decay of dark
      matter particles with $M_X=10^{12}$~GeV clustered in the
      galactic halo.}
     \label{fig:all}
  \end{center}
\end{figure}

We compare in~Fig.~\ref{fig:all} the spectra of baryons, neutrinos and
LSPs expected in the case that UHECRs are produced by the slow decay
of a population of superheavy dark matter particles~$X$ clustered in
the galactic halo. We take as dark matter particle mass
$M_X=10^{12}$~GeV and as scale at which SUSY switches on
$M_\SUSY=400$~GeV. The baryon and neutrino curves are obtained from
Ref.~\cite{SarkarToldra}. The neutralino spectrum shown corresponds to
benchmark point C in Ref.~\cite{Battaglia}. The final neutralino
spectrum does not depend significantly on which benchmark point we
choose, since the dominant contribution in the region of interest
is given by the bino component produced directly by $X$.

\section{Conclusions}

We have used the DGLAP formalism to calculate the spectrum of
neutralinos produced by the decay of particles with mass
$M_X>10^{12}$~GeV. We have concentrated on the electroweak radiation
of neutralinos by partons. This contribution has to be added to the
on-shell contribution calculated with Montecarlo generators. We find
that at large $x$ the radiative contribution is slightly larger.

For $x>0.3$, LSPs dominate over baryons, photons and neutrinos and
their flux on Earth is non-negligible. If UHECR are produced by the
decay of superheavy particles and SUSY is a low energy symmetry of
nature, then forthcoming observatories sensitive to weakly interacting
UHECRs should detect a flux of ultra high energy neutralinos.

\section*{Addendum}
As this manuscript was being finished, Ref.\cite{BarbotDrees} appeared. This
paper also discusses neutralino production by the decay of super-heavy 
dark matter particles using DGLAP evolution.

\section*{Acknowledgements}
We would like to thank Venya Berezinsky, Celine Boehm, Graham Ross and
specially Subir Sarkar for discussion. A.I. would like to thank PPARC
for financial support. R.T. was supported by a Marie Curie Fellowship
No.~HPMF-CT-1999-00268 and by the Leverhulme Trust.


\begin{thebibliography}{99}

\bibitem{NaganoWatson}
M. Nagano, A.A. Watson, Rev. Mod. Phys. 72 (2000) 689.
%%CITATION = RMPHA,72,689;%%

\bibitem{Hill}
C.T. Hill, Nucl. Phys. B 224 (1983) 469.
%%CITATION = NUPHA,B224,469;%%

\bibitem{td}
C.T. Hill, D.N. Schramm, T.P. Walker, Phys. Rev. D 36 (1987) 1007;\\
%%CITATION = PHRVA,D36,1007;%%
P. Bhattacharjee, C.T. Hill, D.N. Schramm, Phys. Rev. Lett. 69 (1992) 567;\\
%%CITATION = PRLTA,69,567;%%
P. Bhattacharjee, G. Sigl, Phys. Rev. D 51 (1995) 4079;\\
%%CITATION = ASTRO-PH 9412053;%%
V. Berezinsky, A. Vilenkin, Phys. Rev. Lett. 79 (1997) 5202.
%%CITATION = ASTRO-PH 9704257;%%

\bibitem{bbv98}
V. Berezinsky, P. Blasi, A. Vilenkin, Phys. Rev. D 58 (1998) 103515.
%%CITATION = PHRVA,D58,103515;%%

\bibitem{tdgamnu}
 G. Sigl, S. Lee, P. Bhattacharjee, S. Yoshida, Phys.
  Rev. D 59 (1999) 043504.
%%CITATION = HEP-PH 9809242;%%

\bibitem{chor}
V.A. Kuzmin, V.A. Rubakov, Phys. Atom. Nucl. 61 (1998) 1028.
%%CITATION = ASTRO-PH 9709187;%%

\bibitem{bkv97}
V. Berezinsky, M. Kachelrie{\ss}, A. Vilenkin, 
 Phys. Rev. Lett. 79 (1997) 4302.
%%CITATION = ASTRO-PH 9708217;%%

\bibitem{BirkelSarkar}
M. Birkel, S. Sarkar, Astropart. Phys. 9 (1998) 297.
%%CITATION = HEP-PH 9804285;%%

\bibitem{gravprod}
D. Chung, E.W. Kolb, A. Riotto, Phys. Rev. D 59 (1999) 023501;\\
%%CITATION = HEP-PH 9802238;%%
V. Kuzmin, I. Tkachev,  Phys. Rev. D 59 (1999) 123006;\\
%%CITATION = HEP-PH 9809547;%%
D. Chung, P. Crotty, E.W. Kolb, A. Riotto, Phys. Rev. D 64 (2001) 043503.
%%CITATION = HEP-PH 0104100;%%

\bibitem{reheatprod} 
D.J. Chung, E.W. Kolb, A. Riotto, Phys. Rev. D 60 (1999) 063504.
%%CITATION = HEP-PH 9809453;%%

\bibitem{BerezKachel01}
V. Berezinsky and M. Kachelrie{\ss}, Phys. Rev. D 63 (2001) 034007.
%%CITATION = HEP-PH 0009053;%%

\bibitem{bk98}
V. Berezinsky, M. Kachelrie{\ss}, Phys. Lett. B 434 (1998) 61.
%%CITATION = HEP-PH 9803500;%%

\bibitem{Rubin} 
N. Rubin, M. Phil. Thesis, Cavendish Laboratory, 
University of Cambridge (1999) 
{\tt (http://www.stanford.edu/\~{ }nrubin/Thesis.ps)}.

\bibitem{FodorKatz} 
Z. Fodor and S.D. Katz, Phys. Rev. Lett. 86 (2001) 3224.
%%CITATION = HEP-PH 0008204;%%

\bibitem{SarkarToldra}
S. Sarkar and R. Toldr\`a, Nucl. Phys. B 621 (2002) 495.
%%CITATION = HEP-PH 0108098;%%

\bibitem{Toldra}
R.~Toldr\`a, Comput. Phys. Commun. 143 (2002) 287.
%%CITATION = HEP-PH 0108127;%%

\bibitem{ChungFarrarK}
D.J. Chung, G.R. Farrar and E.W. Kolb,
Phys.\ Rev.\ D 57 (1998) 4606.
%%CITATION = ASTRO-PH 9707036;%%

\bibitem{LSP}
J.~R.~Ellis, T.~Falk, G.~Ganis and K.~A.~Olive,
Phys. Rev. D 62 (2000) 075010.
%%CITATION = HEP-PH 0004169;%%

\bibitem{AveHinton}
M. Ave, J.A. Hinton, R.A. V{\'a}zquez, A.A. Watson, E. Zas,
 Phys. Rev. Lett. 85 (2000)~2244.
%%CITATION = ASTRO-PH 0007386;%%

\bibitem{HalzenHooper}
F.~Halzen and D.~Hooper,
hep-ph/0110201.
%%CITATION = HEP-PH 0110201;%%

\bibitem{Sigl}
G. Sigl, hep-ph/0109202.
%%CITATION = HEP-PH 0109202;%%

\bibitem{BerezKachel98}
V.~Berezinsky and M.~Kachelrie{\ss},
Phys. Lett. B 422 (1998) 163.
%%CITATION = HEP-PH 9709485;%%

\bibitem{AltarelliParisi} 
G. Altarelli and G. Parisi, Nucl. Phys. B 126 (1977) 298.
%%CITATION = NUPHA,B126,298;%%

\bibitem{DGL} 
L.N. Lipatov, Sov. J. Nucl. Phys. 20 (1975) 94;\\
%%CITATION = SJNCA,20,94;%%
V.N. Gribov and L.N. Lipatov, Sov. J. Nucl. Phys. 15 (1972) 438;\\
%%CITATION = YAFIA,15,781;%%
Yu.L. Dokshitzer, Sov. Phys. JETP 46 (1977) 641.
%%CITATION = SPHJA,46,641;%%

\bibitem{KounnasRoss} 
C. Kounnas and D.A. Ross, Nucl. Phys. B 214 (1983) 317. 
%%CITATION = NUPHA,B214,317;%%

\bibitem{JonesLlewellyn} 
S.K. Jones and C.H. Llewellyn-Smith, Nucl. Phys. B 217 (1983) 145.
%%CITATION = NUPHA,B217,145;%%

\bibitem{DeWitt}
R.J.~DeWitt, L.M.~Jones, J.D.~Sullivan, D.E.~Willen and H.W.~Wyld,
Phys. Rev. D 19 (1979) 2046
[Erratum-ibid. D 20 (1979) 1751].
%%CITATION = PHRVA,D19,2046;%%

\bibitem{Nicolaidis}
A.~Nicolaidis,
Nucl. Phys. B 163 (1980) 156.
%%CITATION = NUPHA,B163,156;%%

\bibitem{FurmanskiPetronzio} 
W. Furma\'{n}ski and R. Petronzio, Nucl. Phys. B 195 (1982) 237.
%%CITATION = NUPHA,B195,237;%%

\bibitem{Battaglia}
M.~Battaglia {\it et al.},
Eur. Phys. J. C 22 (2001) 535.
%%CITATION = HEP-PH 0106204;%%

\bibitem{BarbotDrees}
C.~Barbot and M.~Drees,
hep-ph/0202072.
%%CITATION =  HEP-PH 0202072;%%

\end{thebibliography}
\end{document}